\documentclass[usenatbib]{mn2e}
\usepackage{txfonts}
\usepackage{amssymb}
\usepackage{wasysym}
\usepackage{graphicx}

\title{Virial theorem for radiating accretion discs}
\author[Patryk Mach]{Patryk Mach\\
M. Smoluchowski Institute of Physics, Jagiellonian University Reymonta 4, 30-059 Krak\'ow, Poland}

\date{}

\begin{document}
\maketitle
\label{firstpage}

\begin{abstract}
A continuum version of the virial theorem is derived for a radiating self-gravitating accretion disc around a compact object. The central object is point-like, but we can avoid the regularization of its gravitational potential. This is achieved by applying a modified Pohozaev--Rellich identity to the gravitational potential of the disk only. The theorem holds for general stationary configurations, including discontinuous flows (shock waves, contact discontinuities). It is used to test numerical solutions of a model of self-gravitating radiative accretion discs. The presented virial theorem should be useful in the analysis of those (possibly radiating) hydrodynamical systems in astrophysics where the central mass and the mass of the fluid are comparable and none of them can be neglected.
\end{abstract}


\section{Introduction}

The standard virial theorem applies to a stationary configuration of self-gravitating fluid (a star). A textbook level exposition of its Newtonian version can be found in \citet{tassoul} or \citet{collins}. The general-relativistic formulation is still unsatisfactory, although preliminary studies are done in \citet{gourgoulhon} and \citet{karkowski_malec}. In this paper we go back to Newtonian gravity, and formulate the virial theorem for a steady radiative accretion disc around a central point mass. 

This result emerged from an investigation of stationary radiative accretion discs done in \citet{mach_malec}, where a test of numerical solutions was required. The choice of the virial theorem was natural---one of important applications of its standard version is to test numerical models of rotating stars \citep[see e.g.][]{hachisu}.

The radiation was included in the virial theorem formulated in \citet{anand}, but under too stringent assumptions on the radiation stress tensor. In this paper the radiation stress tensor appears in the equation expressing the conservation of the momentum, but there is no need to specify its form.

The important feature of our work, absent in the former investigations, is the presence of the potential due to a point-like central mass. It can cause difficulties in the standard derivation, since meaningless (singular) terms do appear. This point requires an explanation. The classic Newtonian version of the virial theorem can be written as
\[ \int_{\mathbb R^3} d^3 x \, \frac{1}{2} \rho \Phi + 2 \int_{\mathbb R^3} d^3 x \, \frac{1}{2} \rho | \mathbf U |^2 + 2 \int_{\mathbb R^3} d^3 x \, \frac{3}{2} p = 0, \]
where $\Phi$ is the gravitational potential, $\rho$ the mass density, $\mathbf U$ the velocity of the fluid, and $p$ its pressure. Imagine that we try to apply this version to a system consisting of a stationary toroid of fluid and a point mass $M_c$ located at $\mathbf x = 0$. Naively, we could write $\rho = \rho_\mathrm{f} + M_\mathrm{c} \delta^{(3)}(\mathbf x)$, where $\rho_\mathrm{f}$ denotes the density of the fluid only. The corresponding gravitational potential would be $\Phi = \Phi_\mathrm{f} + \Phi_\mathrm{Kep}$. Here $\Phi_\mathrm{f}$ denotes the gravitational potential of the fluid, and $\Phi_\mathrm{Kep} = - G M_\mathrm{c}/|\mathbf x|$ is the Keplerian potential of the point mass ($G$ denotes the gravitational constant). The fluid related quantities satisfy $\Delta \Phi_\mathrm{f} = 4 \pi G \rho_\mathrm{f}$, while for the central mass we have $\Delta \Phi_\mathrm{Kep} = 4 \pi G M_\mathrm{c} \delta^{(3)}(\mathbf x)$. Clearly $\Delta \Phi = 4 \pi G \rho$, as desired. The integral
\begin{eqnarray}
\label{naive}
\int_{\mathbb R^3} d^3 x \, \rho \Phi & = & \int_{\mathbb R^3} d^3 x \, \rho_\mathrm{f} \Phi_\mathrm{f} + \int_{\mathbb R^3} d^3 x \, \rho_\mathrm{f} \Phi_\mathrm{Kep} \\
&+& \int_{\mathbb R^3} d^3 x \, M_\mathrm{c} \delta^{(3)}(\mathbf x) \Phi_\mathrm{f} + \int_{\mathbb R^3} d^3 x \, M_\mathrm{c} \delta^{(3)}(\mathbf x)\Phi_\mathrm{Kep} \nonumber
\end{eqnarray}
is, however, meaningless due to the last term on the right-hand side. In the following we will show that this term does not appear in the proper calculation.

The virial theorem of this paper is presented in the next section. It is then used in order to test the numerical model of a radiative, self-gravitating disc of gas accreting onto a compact object derived in \citet{mach_malec}. The details of this model are given in Section 3.1, and the virial check is presented in Section 3.2.

\section{The theorem}

Consider a steady self-gravitating configuration of fluid interacting with radiation. It satisfies the Euler equation:
\begin{equation}
\label{euler_eq}
\nabla \cdot \left( \rho_\mathrm{f} \mathbf U \otimes \mathbf U + p + \mathcal P \right) = - \rho_\mathrm{f} \nabla \Phi,
\end{equation}
which we deliberately write in the conservative form \citep[see e.g.][]{castor}. Here, as in Section 1, $\mathbf U$ is the velocity, $\rho_\mathrm{f}$ the density and $p$ the pressure of the fluid. The gravitational potential is denoted by $\Phi$, and the radiation is described by the radiation pressure tensor:
\[ \mathcal P = \frac{1}{c} \int_0^\infty d \nu \int_{4 \pi} \hat{\mathbf k} \otimes \hat{\mathbf k} I_\nu \, d \Omega, \]
where $I_\nu$ is the radiation intensity, and $\hat{\mathbf k}$ denotes a unit vector oriented in the direction in which the propagation of radiation is considered. As usual, $\nu$ denotes the radiation frequency, and $c$ is the speed of light. Here and in what follows, the divergence of any symmetric tensor $\mathcal F$ is understood as a vector field field with components $(\nabla \cdot \mathcal F)^j = \partial_i \mathcal F^{ij}$ in Cartesian coordinates. The symbol $\mathbf u \otimes \mathbf v$ is used to denote the dyadic product of two vector fields $\mathbf u$ and $\mathbf v$. It is defined as a tensor with components $(\mathbf u \otimes \mathbf v)^{ij} = u^i v^j$.

We assume that the gas is orbiting around a point mass $M_\mathrm{c}$, and that the gas and the point mass are isolated from each other (i.e., the mass is located outside of the support of $\rho_\mathrm{f}$). We fix the origin of the coordinate system so that it coincides with the position of the mass $M_\mathrm{c}$. We will work in Cartesian coordinates in this section; they are denoted by $x^i$, with $\mathbf x = (x^1, x^2, x^3)$.

The gravitational potential $\Phi$ can be expressed as $\Phi = \Phi_\mathrm{Kep} + \Phi_\mathrm{f}$. Here the Keplerian potential $\Phi_\mathrm{Kep}$ is given by $\Phi_\mathrm{Kep} = - G M_\mathrm{c}/|\mathbf x|$, and the gravitational potential of the fluid satisfies the Poisson equation:
\begin{equation}
\label{poisson_eq}
\Delta \Phi_\mathrm{f} = 4 \pi G \rho_\mathrm{f}.
\end{equation}

We will now use a modified version of the Pohozaev--Rellich identity \citep{pohozaev, rellich} for the above equation. Consider a vector field
\[ \mathbf w = \left( (\mathbf x \cdot \nabla) \Phi_\mathrm{f} + \Phi_\mathrm{f}/2 \right)  \nabla \Phi_\mathrm{f} - \mathbf x \left| \nabla \Phi_\mathrm{f} \right|^2/2 + 4 \pi G p \mathbf x. \]
A straightforward calculation making use of the Poisson equation (\ref{poisson_eq}) shows that
\[ \nabla \cdot \mathbf w = 4 \pi G \left( \rho_\mathrm{f} \Phi_\mathrm{f}/2 + \rho_\mathrm{f} \mathbf x \cdot \nabla \Phi_\mathrm{f} + \mathbf x \cdot \nabla p + 3 p \right). \]
Note that $\Phi_\mathrm{Kep}$ does not enter the above formula at this stage, and there are no aforementioned issues with the singularity at $\mathbf x = 0$. It follows from the Euler equation (\ref{euler_eq}) that
\begin{eqnarray*}
\rho_\mathrm{f} \mathbf x \cdot \nabla \Phi_\mathrm{f} + \mathbf x \cdot \nabla p  & = & - \rho_\mathrm{f} \mathbf x \cdot \nabla \Phi_\mathrm{Kep} - \mathbf x \cdot \left( \nabla \cdot (\rho_\mathrm{f} \mathbf U \otimes \mathbf U) \right) \\
&&- \mathbf x \cdot (\nabla \cdot \mathcal P).
\end{eqnarray*}
Moreover, for the Keplerian potential, we have $- \mathbf x \cdot \nabla \Phi_\mathrm{Kep} = \Phi_\mathrm{Kep}$. Thus
\begin{eqnarray*}
\nabla \cdot \mathbf w & = & 4 \pi G \rho_\mathrm{f} \left((\Phi + \Phi_\mathrm{Kep})/2 - \mathbf x \cdot \left( \nabla \cdot (\rho_\mathrm{f} \mathbf U \otimes \mathbf U) \right) + 3 p \right. \\
&& \left. - \mathbf x \cdot (\nabla \cdot \mathcal P)  \right).
\end{eqnarray*}

We will assume further that the fluid is finite, i.e. the support of the density $\rho_\mathrm{f}$ and the pressure $p$ is compact. The reader can consult \cite{mach_simon} for the necessary assumptions in the infinite case. We integrate the above equation over a ball $\mathcal B(0,R)$ of radius $R$, centered at $\mathbf x = 0$ and containing the support of the pressure and the density. We then take the limit of $R \to \infty$. The left hand side of the obtained equation can be converted into a surface term that vanishes, since the gravitational potential $\Phi_\mathrm{f}$ falls off like $1/|\mathbf x|$ asymptotically. Moreover, the second term on the right-hand side can be integrated by parts. This gives the main result of this paper:
\begin{eqnarray}
0 & = & 4 \pi G \int_{\mathbb{R}^3} d^3 x \, \left( \rho_\mathrm{f} (\Phi + \Phi_\mathrm{Kep})/2 + \rho_\mathrm{f} | \mathbf U |^2 + 3 p - \mathbf x \cdot (\nabla \cdot \mathcal P) \nonumber \right) \\
& = & 4 \pi G \left( E_\mathrm{pot} + 2 E_\mathrm{kin} + 2 E_\mathrm{therm} + \tilde E  \right). \label{virial_th}
\end{eqnarray}
Here $E_\mathrm{pot} = \int_{\mathbb R^3} d^3 x \, \rho_\mathrm{f} (\Phi + \Phi_\mathrm{Kep})/2$ is the total potential energy,  $E_\mathrm{kin} = \int_{\mathbb R^3} d^3 x \, \rho_\mathrm{f} | \mathbf U |^2/2$ denotes the bulk kinetic energy, $E_\mathrm{therm} = \int_{\mathbb R^3} d^3 x \, 3 p / 2$ is the internal thermal energy, and $\tilde E = - \int_{\mathbb R^3} d^3 x \, \mathbf x \cdot (\nabla \cdot \mathcal P)$ describes the interaction with the radiation.

The potential $\Phi_\mathrm{Kep}$ enters the above expression twice: once as a part of $\Phi$, and then explicitly. This corresponds exactly to the first three integrals appearing on the right-side of equation (\ref{naive}), as anticipated. To see this note that the second and third integrals on the right-hand side of equation (\ref{naive}) are equal. Indeed,
\begin{eqnarray*}
\int_{\mathbb R^3} d^3 x \, M_\mathrm{c} \delta^{(3)}(\mathbf x) \Phi_\mathrm{f} & = & \int_{\mathbb R^3} d^3 x \, \Delta \Phi_\mathrm{Kep} \Phi_\mathrm{f}/(4 \pi G) \\
& = & \int_{\mathbb R^3} d^3 x \, \Phi_\mathrm{Kep} \Delta \Phi_\mathrm{f}/(4 \pi G) \\
& = & \int_{\mathbb R^3} d^3 x \, \Phi_\mathrm{Kep} \rho_\mathrm{f}.
\end{eqnarray*}
Here the surface terms vanish because of the asymptotic fall-off of both gravitational potentials.

Also note that the term $\tilde E = - \int d^3 x \, \mathbf x \cdot (\nabla \cdot \mathcal P)$ can be integrated by parts yielding
\[ \tilde E = - \int d^3 x \, \mathbf x \cdot (\nabla \cdot \mathcal P) = \int d^3 x \, E + \mathrm{surface \; term},  \]
where
\[ E = \mathrm{Tr} \, \mathcal P = \frac{1}{c} \int_0^\infty d \nu \int_{4 \pi} I_\nu \, d\Omega \]
is the radiation energy density. In general, the surface term appearing here does not vanish for radiating systems. We will not be using this form in the following sections.

The derivatives appearing in the above formulae can be understood in a weak (integral) sense (to see this, one has to analyse all steps of the derivation of the theorem with respect to the character of derivatives). This fact is especially important for hydrodynamical quantities $p$, $\rho_\mathrm{f}$, $\mathbf U$ that can be discontinuous. The gravitational potential $\Phi_\mathrm{f}$ is assumed to be smooth, and its derivatives appearing in the calculation can be understood in the classical sense. The obtained result is thus quite general. Contact discontinuities and stationary shock waves are allowed in the underlying solution of the Poisson--Euler system of equations (\ref{euler_eq}--\ref{poisson_eq}). An introduction into the notion of weak or integral solutions of nonlinear partial differential equations with a special emphasis put on conservation laws such as (\ref{euler_eq}) can be found in \citet{evans}.

\section{Application}
\subsection{Accretion model}

We will demonstrate that the above theorem allows for an easy check of the validity of self-gravitating steady accretion discs with radiation obtained numerically in \citet{mach_malec}.

In the following we sketch the construction of the model; details can be found in \citet{mach_malec}. Consider a quasi-stationary configuration of perfect fluid accreting onto the central point mass $M_\mathrm{c}$.  Assume that there exist some mechanism that produces radiation with the emissivity per unit mass $j$, and that the radiation interacts with the gas via Thompson scattering. It can be shown that the radiative flux
\[ \mathbf F = \int_0^\infty d \nu \int_{4 \pi} \hat{\mathbf k} I_\nu \, d \Omega \]
satisfies $\nabla \cdot \mathbf F = \rho_\mathrm{f} j$. Moreover, the momentum exchange rate appearing in (\ref{euler_eq}) is $\nabla \cdot \mathcal P = - (\rho_\mathrm{f} \kappa/c) \mathbf F$, where $\kappa$ is the scattering opacity. The energy conservation equation can be written as
\begin{equation}
\label{energy_eq}
\nabla \cdot \left(\rho_\mathrm{f} \mathbf U \left( h + | \mathbf U |^2/2 + \Phi \right) + \mathbf F \right) = 0,
\end{equation}
where $h$ denotes the specific enthalpy.

We will now restrict ourselves to polytropic equations of state $p = K \rho_\mathrm{f}^\Gamma$ and simple axisymmetric flows. Let $(r, \phi, z)$ denote cylindrical coordinates. Assume the velocity of the fluid in the form $\mathbf U = U \partial_r + \omega \partial_\phi$, with $U \ll r \omega$. Define the mass accretion rate function $\dot M = -2 \pi U r \rho_\mathrm{f}$. It follows from the continuity equation
\[ \nabla \cdot (\rho_\mathrm{f} \mathbf U) = 0, \]
that $\dot M = \dot M(z)$. It is then shown that $\mathbf F$ can be expressed as a gradient of a potential: $(\kappa/c) \mathbf F = \nabla \Psi$.

The solution of the model is computed from the following equations: the integrated Euler equation (\ref{euler_eq})
\begin{equation}
\label{int_euler_eq}
h + \Phi_\mathrm{Kep} + \Phi_\mathrm{f} + \Phi_\mathrm{c} - \Psi = C,
\end{equation}
where $C$ denotes an integration constant, the energy equation (\ref{energy_eq}) rewritten as
\begin{equation}
\label{psi_eq}
\Delta \Psi = \frac{\kappa \dot M}{2 \pi c r} \left(\partial_r \Psi + 2 r \omega^2 + r^2 \omega \partial_r \omega \right),
\end{equation}
and the Poisson equation for the gravitational potential (\ref{poisson_eq}). Here, as usual, the rotation law $\omega = \omega(r)$ has to be specified a priori, and the centrifugal potential $\Phi_\mathrm{c}$ is defined as
\[ \Phi_\mathrm{c} = - \int^r dr^\prime \, r^\prime \omega^2(r^\prime). \]

The radiation mechanism (the emissivity per unit mass $j$) is not specified at the beginning. Instead, we prescribe the accretion rate function $\dot M$ and require stationarity of the resulting configuration. The complete solution can be then computed form the energy and momentum conservation equations, together with the emissivity $j$. In this way we are following the spirit of classic papers on astrophysical accretion. It has been pointed out by old masters \citep[see e.g.][]{pringle} that once a steady accretion occurs, one can eliminate the viscosity from the description. The effective model needs only $\dot M$; no specific information on the production of radiation is required.

As a consistency check, it remains to be verified that the condition $U \ll r \omega$ is valid in the obtained solution. This requirement concurs with the fact that solutions of the model do only exist for small luminosities, and thus for small $\dot M$.

Quasi-stationarity of the model is understood in the usual sense; some matter must be delivered to the system from an outside reservoir at a small rate $\dot M$. The central mass $M_\mathrm{c}$ grows in time, but this growth should be negligible.

Equations (\ref{poisson_eq}), (\ref{int_euler_eq}) and (\ref{psi_eq}) have a scaling symmetry that can be used to convert them into a dimensionless form. Assume that the disc spreads up to $r = r_\mathrm{out}$ at the equatorial plane, and the maximal density of the gas is $\rho_\mathrm{max}$. The quantity $u = G R^2_\mathrm{out} \rho_\mathrm{max}$ has the dimension of the potentials. It can be used to define $\tilde h = h/u$, $\tilde \Phi_\mathrm{Kep} = \Phi_\mathrm{Kep}/u$, $\tilde \Phi_\mathrm{f} = \Phi_\mathrm{f}/u$, $\tilde \Phi_\mathrm{c} = \Phi_\mathrm{c}/u$, $\tilde \Psi = \Psi/u$, and $\tilde \omega = \omega r_\mathrm{out}/\sqrt{u}$. Spatial dimensions and the density can be also rescaled: $\tilde {\mathbf x} = \mathbf x/r_\mathrm{out}$ and $\tilde \rho_\mathrm{f} = \rho / \rho_\mathrm{max}$. The potential $\tilde \Phi_\mathrm{Kep}$ is then expressed as $\tilde \Phi_\mathrm{Kep} = - \tilde M_\mathrm{c}/|\tilde {\mathbf x}|$, where $\tilde M_\mathrm{c} = M_\mathrm{c}/(\rho_\mathrm{max} r_\mathrm{out}^3)$. Equations (\ref{poisson_eq}), (\ref{int_euler_eq}) and (\ref{psi_eq}) can be now reduced to
\begin{equation}
\label{rescaled_euler}
\tilde h + \tilde \Phi_\mathrm{Kep} + \tilde \Phi_\mathrm{f} + \tilde \Phi_\mathrm{c} - \tilde \Psi = C,
\end{equation}
\[ \tilde \Delta \tilde \Phi_\mathrm{f} = 4 \pi \tilde \rho, \]
\[ \tilde \Delta \tilde \Psi = \frac{\kappa \dot M}{2 \pi c \tilde r} \left( \partial_{\tilde r} \tilde \Psi + 2 \tilde r \tilde \omega^2 + \tilde r^2 \tilde \omega \partial_{\tilde r} \tilde \omega \right), \]
where $\tilde \Delta$ denotes the laplacian with respect to the rescaled coordinates $\tilde {\mathbf x}$.

\subsection{Numerical tests}

\begin{figure}
\begin{center}
\includegraphics[width=8cm]{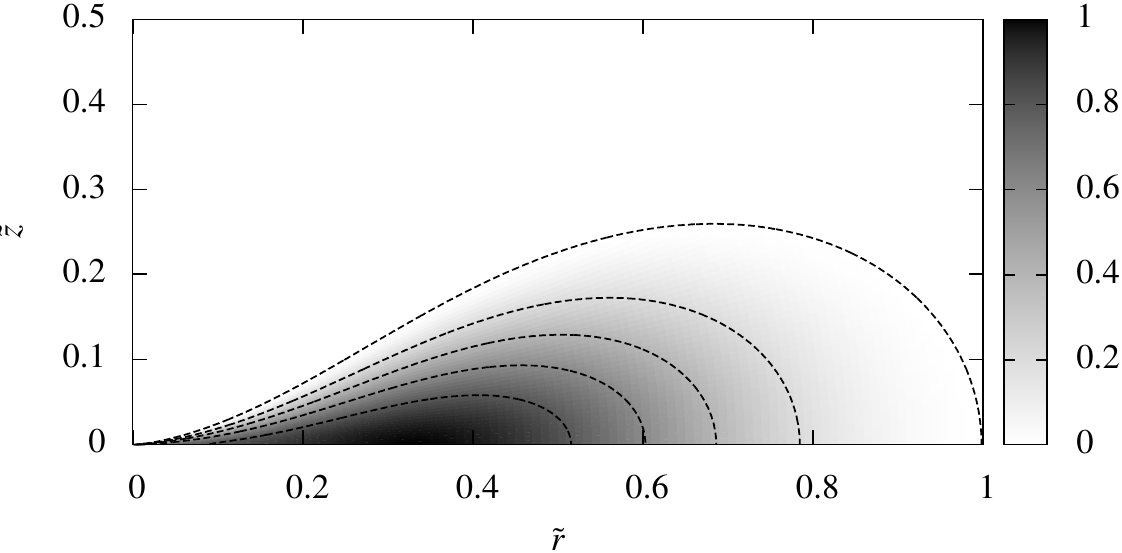}
\end{center}
\caption{The density plot of the radiating accretion disc solution. Due to the assumed symmetry, only a section through a meridian upper half-plane is shown. The rescaled density $\tilde \rho$ is coded in grey-scale.}
\end{figure}

Numerical solutions of the model can be found by extending the classic Self-Consistent-Filed (SCF) method that was initially used to model equilibrium configurations of rotating fluids (stars) \citep{ostriker_mark}. The basic idea of the SCF method is to convert the Poisson equation for the gravitational potential into the integral formula, regularise the Poisson kernel by the expansion in the Legendre polynomials, and iterate it together with the integrated Euler equation of the form (\ref{rescaled_euler}) until some level of convergence is reached. In the case of the accretion model presented here, there is one more Poisson-like equation to be solved, but the overall scheme can be retained. Details of this numerical method can be found in \citet{mach_malec}.

As an illustration of the virial theorem of this paper we consider a numerical solution obtained for the rotation law $\tilde \omega(\tilde r) = \bar \omega / \tilde r^{3/2}$, where $\bar \omega = \mathrm{const}$. We search for a disk solution with $\tilde M_\mathrm{c} = 1$ and $r_\mathrm{in}/r_\mathrm{out} = 10^{-3}$, where $r_\mathrm{in}$ denotes the inner equatorial radius of the disk.  We assume the polytropic exponent $\Gamma = 5/3$ and the accretion rate function
\[ \kappa \dot M /c = 10^{-3} \exp \left( - (50 \tilde z)^2 \right) \]
(note that $\kappa \dot M /c$ is a dimensionless quantity).

The obtained distribution of $\tilde \rho$ is shown on Fig.~1. The mass of the disk equals $M_\mathrm{f} = \int_{\mathbb{R}^3}d^3 x \, \rho_\mathrm{f} \approx 0.294 M_\mathrm{c}$. Choosing $r_\mathrm{out} = 10 \, \mathrm{pc}$ and $\rho_\mathrm{max} = 10^{-17} \mathrm{g \, cm^{-3}}$ we get $M_\mathrm{c} \approx 1.48 \times 10^8 M_{\astrosun}$. The mass of the disk is then $M_\mathrm{f} \approx 4.34 \times 10^7 M_{\astrosun}$ (solar masses). The total luminosity can be defined as $L = \int d \mathbf S \cdot \mathbf F$, where the integral is evaluated over the disk surface, and $d \mathbf S$ denotes an outward oriented surface element. It yields $L \approx 2.06 \times 10^9 L_{\astrosun}$, where $L_{\astrosun}$ is the solar luminosity. Such a soluion could serve as a model of an accretion disc around an ultramassive galcatic black hole.

For the same solution we get $E_\mathrm{kin}/|E_\mathrm{pot}| \approx 0.446$, $E_\mathrm{term}/|E_\mathrm{pot}| \approx 5.33 \times 10^{-2}$, and $\tilde E/|E_\mathrm{pot}| \approx 3.74 \times 10^{-4}$. Also $\left(\int_{\mathbb R^3} d^3 x \, \rho_\mathrm{f} \Phi/2 \right)/E_\mathrm{pot} \approx 0.559$, and $\left(\int_{\mathbb R^3} d^3 x \, \rho_\mathrm{f} \Phi_\mathrm{Kep}/2\right)/E_\mathrm{pot} \approx 0.441$.

The virial check can be performed by computing $\epsilon = |E_\mathrm{pot} + 2 E_\mathrm{kin} + 2 E_\mathrm{therm} + \tilde E|/|E_\mathrm{pot}|$, according to equation (\ref{virial_th}). In our case $\epsilon \approx 10^{-8}$ for the solution obtained on the numerical grid with the resolution of $5000 \times 5000$ points and the multipole expansion using more that 20 Legenrde polynomials. Note that this value is quite a few orders of magnitude smaller than each of its constituent terms. This confirms the validity of the numerical solution. It is also an independent check of the absence of trivial mistakes in the derivation of the virial theorem of this paper.

\section{Final remarks}

We shall warn the reader that a set of functions satisfying the virial theorem does not have to be a solution of the Poisson--Euler system of equations (\ref{euler_eq}--\ref{poisson_eq}). \citet{odrzywolek} presented an iterative analytical scheme for obtaining such solutions, where in each step the virial theorem was satisfied, although a convergence to the solution was not yet reached.

Despite this, the virial check still belongs to the very few tools that allow one to verify the correctness of numerical solutions of the Poisson--Euler system  of equations (\ref{euler_eq}--\ref{poisson_eq}). This paper provides such a test for a stationary accretion system of self-gravitating gas.

The virial theorem presented here can be used in the analysis of those stationary hydrodynamical systems, where the central mass and the mass of the fluid are comparable, and thus none of them can be neglected. Models of such systems can be found in \citet{hashimoto} or \citet{sobolev}.

Magnetic fields can be included in the analysis in the manner similar to the radiation. In the astrophysical context this was done by \citet{chandrasekhar}. As a result, an additional term representing the magnetic energy appears in (\ref{virial_th}).

\section*{Acknowledgements}

The author would like to thank Edward Malec and Andrzej Odrzywo\l ek for their comments on the draft of this paper.

The research was carried out with the supercomputer `Deszno' purchased thanks to the financial support of the European Regional Development Fund in the framework of the Polish Innovation Economy Operational Program (contract no. POIG.02.01.00-12-023/08).

\end{document}